\documentclass[published,cits]{JINST}
\usepackage{microtype}
\usepackage{wasysym}
\fussy

\title{Progress on large area GEMs}
\author{Serge Duarte Pinto$^{ab}$\thanks{Corresponding author: Serge.Duarte.Pinto@cern.ch}, Marco Villa$^{ab}$\thanks{Marco.Villa@cern.ch}, Matteo Alfonsi$^a$, Ian Brock$^b$, Gabriele Croci$^a$, Eric David$^a$, Rui de Oliveira$^a$, Leszek Ropelewski$^a$, Miranda van Stenis$^a$\\
\llap{$^a$}CERN,\\ Geneva, Switzerland.\\
\llap{$^b$}Physikalisches Institut der Universit\"at Bonn,\\ Bonn, Germany.\\
}

\abstract{
\ In 2008, a triple \textsc{gem} detector prototype with an area of $\sim~2000$~cm$^\mathbf{2}$ has been constructed, based on foils of $66 \times 66$ ~cm.
\textsc{Gem}s~of such dimensions had not been made before, and innova\-tions to the existing technology were introduced to build this detector.
This paper discusses these innovations and presents further work on large area \textsc{gem} development.

\qquad A \emph{single-mask technique} overcomes the cumbersome practice of alignment of two masks, which limits the achievable lateral size.
The holes obtained with this technique are conical, and have a so-called \emph{rim}, a small insulating clearance around the hole in the substrate.
Further refinements of this technique allow greater control over the shape of holes and the size of rims.
Also, an improvement in homogeneity over large areas is expected.

\qquad Simulation studies have been done to examine the effect of hole shape on the behavior of \textsc{gem}s.
Such studies can help understanding how to use new enhancements of the technique to optimize performance.

\qquad Many potential applications for large area \textsc{gem}s foresee large production volumes.
Production issues have been studied, and single-mask \textsc{gem}s turn out to be much more suitable for large scale production than standard \textsc{gem}s.
}

\keywords{gaseous detectors, gas electron multipliers, \textsc{gem}, Gaseous imaging and tracking detectors}

\received{August 31, 2009}
\accepted{November 26, 2009}
\published{November 26, 2009}

\begin{document}

\section{Introduction}
\label{intro}
The Gas Electron Multiplier (\textsc{gem}) is a gaseous charge amplification structure, invented in 1998 \cite{firstGEM}.
It has found many applications in high-energy physics and other fields of research.
Most commonly used in a triple \textsc{gem} cascade, it is among the most rate capable gas detectors currently available.
Its spatial resolution of optimally $\sim 50 \mu$m and time resolution of a few nanoseconds makes it suitable for both tracking and triggering applications.
Two of the current \textsc{lhc} experiments (\textsc{totem}~\cite{TOTEM} and \textsc{lhc}b~\cite{LHCb}) use triple \textsc{gem} detectors, both in the forward regions where high rate capability is imperative.
With the plans for a high luminosity upgrade to s\textsc{lhc}, experiments need to prepare for a tenfold increase in rate.

Many of the wire-based gas detectors currently used in the experiments will be unable to cope with these high rates.
The slowly drifting ions cause space charge effects, which limit the rate capability to {10}--{100}~kHz/cm$^2$.
In micropattern gas detectors (\textsc{mpgd}s) rate capability is typically limited by discharge probability rather than space charge effects.
In the case of (cascaded) \textsc{gem}s, the rate capability can be increased by about two orders of magnitude with respect to wire-chambers.
However, \textsc{gem}s with dimensions exceeding $\sim 0.5$~m were not available thus far.

An effort has been started to make \textsc{gem} technology available for such large size detectors~\cite{LargeGEM}.
In 2008, a prototype large area triple \textsc{gem} chamber was made, aimed at an application in the \textsc{totem} experiment, as an upgrade of its T1 tracker~\cite{T1}, currently a cathode-strip wire-chamber.
The active area is $\sim 2000$~cm$^2$ and the \textsc{gem} foils used were {66}~cm $\times$ {66}~cm.
To produce foils of such dimensions, limitations in the standard fabrication procedure had to be overcome.
The standard technique for creating the hole pattern, involving accurate alignment of two masks, was replaced by a \emph{single-mask technique}, see section~\ref{singlemask}, which presents recent developments of this technique.

\section{Single-mask technique}
\label{singlemask}
Production of \textsc{gem} foils is based on the photolithographic processes commonly employed by the printed circuit industry.
\begin{figure}[tb]
\includegraphics[width=\textwidth]{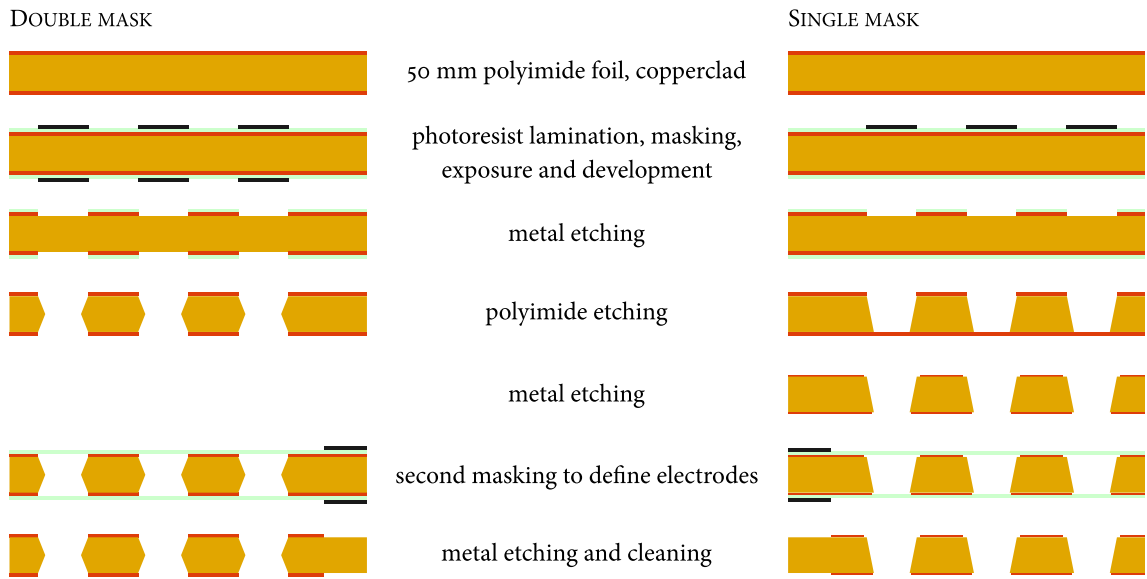}
\caption[Double- vs. single-mask technique]{Schematic comparison of procedures for fabrication of a double-mask \textsc{gem} (left) and a single-mask \textsc{gem} (right).}
\label{DoubleSingle}
\end{figure}
The left hand side of figure~{\ref{DoubleSingle}} shows schematically the steps involved in standard \textsc{gem} fabrication.
The \textsc{gem} hole pattern is transferred by \textsc{uv}-exposure from flexible transparent films to a copper-clad polyimide foil laminated with a photoresistive material.
After development, the foil can be etched in an acid liquid, which removes copper from the holes, but not from where the photoresist still masks the copper.
The next step is the etching of the polyimide substrate, for which the holes in the copper layers act as a mask.

To obtain a homogeneous hole geometry across the foil, it is imperative to keep the alignment error between top and bottom films within 5--{10}~$\mu$m.
As both the films and the base material are flexible this alignment is far from trivial, and when foil dimensions exceed about half a meter this method is hardly feasible anymore.
A way to overcome this hurdle is shown on the right hand side of figure~{\ref{DoubleSingle}}.
By using only one mask to pattern only the top copper layer, no alignment needs to be done.
The bottom copper layer is etched after the polyimide, using the holes in the polyimide as a mask.
The quality and homogeneity of the holes depends now critically on the control of the polyimide etching (which also defines the pattern of the bottom copper layer), where mask alignment used to be the limiting factor.

\subsection{Polyimide etching}
In the development of the single-mask technique, pro\-gress has been made in understanding the crucial parameters and conditions of the polyimide etching process.
Polyimide etching of \textsc{gem}s was traditionally done in a basic aqueous solution containing ethylene diamine and potassium hydroxide (KOH).
The potassium hydroxide has an isotropic etching characteristic, which means that material in contact with the etching liquid is removed at the same rate in all directions.
On the other hand, ethylene diamine etches strongly anisotropically, resulting in wide conical holes.

\begin{figure}[tb]
\center\includegraphics[width=.8\textwidth]{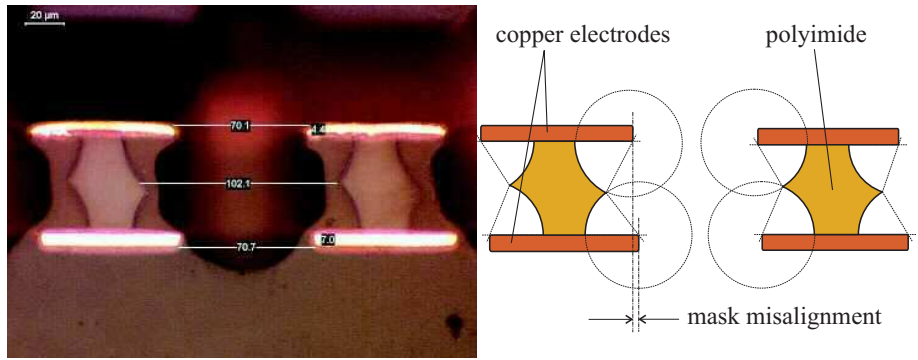}
\caption{Left: cross-section view of isotropically etched polyimide holes (double-mask technique). The etching extends under the copper electrodes in an expanding circular profile, as shown in the reconstruction drawing on the right. \textsc{Gem} foil made by Tech-Etch (Plymouth, \textsc{USA}).}
\label{Tech-Etch}
\end{figure}

Pure isotropic etching results in holes that are always at least twice as wide as they are deep, and polyimide is removed until far under the electrode, as can be seen in figure~{\ref{Tech-Etch}}.
The etching under the copper layer is generally unsatisfactory; the low aspect ratio (defined as the ratio \textit{depth\,/\,width} of a hole) prohibits the use of isotropic etching for a single-mask technique.
When etching with a solution of ethylene diamine only, although no material is removed under the copper, the conical edges are not steep enough to arrive at a high aspect ratio.

In a solution based on both KOH and ethylene diamine, the steepness of the holes can be modified within a certain range by controlling the composition of the liquid.
The tem\-perature must also be well controlled: raising the tem\-perature increases the etching rate of KOH more than that of ethylene diamine.
Furthermore, at temperatures exceeding $\sim${55}$^\circ$C local delaminations between copper and polyimide have been observed, that give rise to irregular etching patterns.
When these parameters are optimized one can make holes of $\sim 90 \mu$m on top, and $\sim 45 \mu$m on the bottom.
Such a profile is used for the large area prototype, see the cross-section in figure~\ref{NoProtection}.
This result does not depend critically on one parameter, and can therefore easily and accurately be reproduced.

More recently, the technique was developed further.
It was found that addition of ethanol to the etching liquid could lead to steeper holes.
Furthermore, the steepness and the level of control over the process can be increased by etching in two phases.
The liquids have different compositions of the same four constituents (ethylene diamine, potassium hydroxide, water, ethanol).

\subsection{Copper etching}
\label{copper}
As shown in figure~{\ref{DoubleSingle}}, the etching of the top electrode using the single-mask technique is done in the same way as with the double-mask technique.
The bottom electrode however, is etched both through the polyimide (which defines the pattern) and from the unprotected outer face of the copper.
This is done by immersion of the foil in the acidic etchant.
If the electrodes are not protected, both electrodes will be etched at the same time, and at the same rate.
Hence the electrodes will be slimmed to half their original thickness or less.
Also, as the copper etching process is isotropic, a clearance arises around the edge of the hole in the polyimide.
This is indicated in figure~{\ref{NoProtection}}.
\FIGURE{\includegraphics[width=\textwidth]{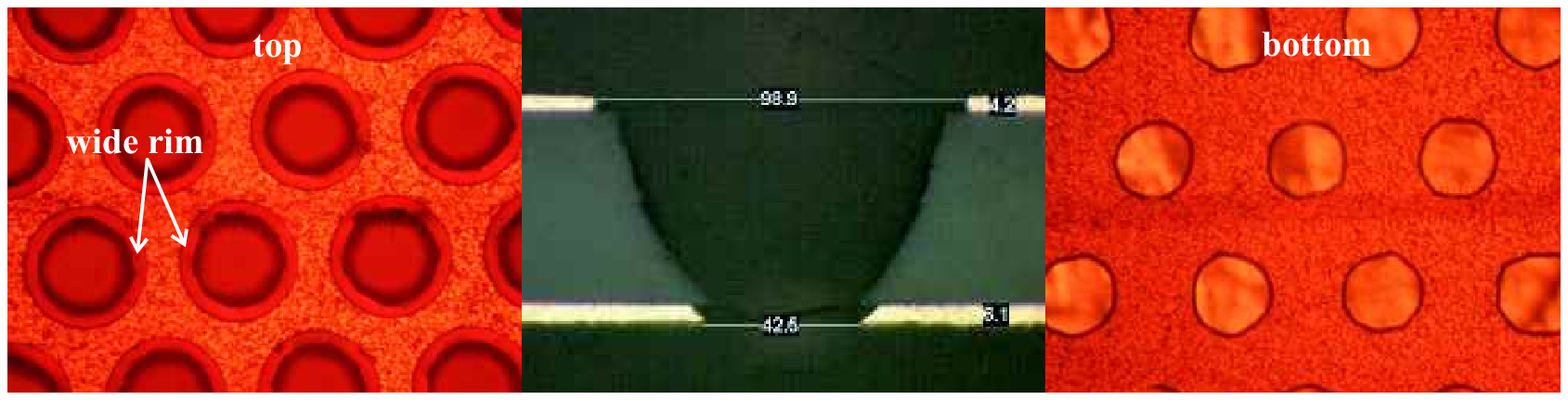}
\caption{Galvanic protection by a very thin gold layer over the top electrode. Due to failing hermeticity of the layer, some copper is etched underneath it.}
\label{NoProtection}}

\FIGURE[l]{\includegraphics[height=29mm]{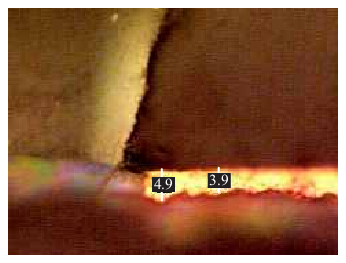}\includegraphics[height=29mm]{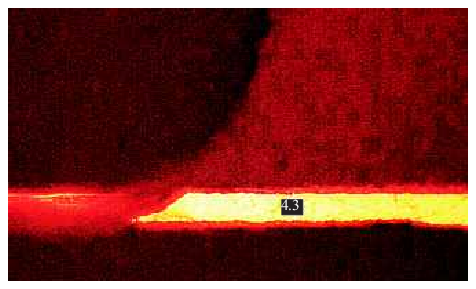}
\caption{Cross section views of copper layers on \textsc{gem}s, at the bottom edge of the hole. Both samples were etched on the surface, to reduce the thickness of copper. The left sample is etched with a microetch, the right sample with chromic acid.}\label{SurfaceEtchants}}
This clearance, usually referred to as the \emph{rim}, is known to degrade time stability of the gain due to dielectric surface charging.
To minimize the width of the rim, the thickness of the electrodes can already be minimized before etching the holes.
Slimming down \textsc{gem} elec\-trodes was shown in \cite{SlimCopper} to leave the detector properties untouched, only reducing the material budget significantly.
In \cite{SlimCopper} a so-called \emph{microetch} liquid based on ammonium persulfate was used.
This type of liquid etches more efficiently at grain boundaries, giving the copper surface a matte look.
Using this surface etchant at large areas, we found that the resulting copper thickness varies by a few microns across the foil.
With the single-mask technique, this causes some variation in hole diameter on the bottom electrodes, which gives rise to gain inhomogeneities.
In addition, the increased surface roughness degrades the definition of the holes in the bottom.
Such rough copper edges around the hole cause discharges in \textsc{gem} operation.

By using chromic acid instead of a microetch as a surface etchant, both issues of inhomogeneity and surface roughness are overcome, leaving a shiny surface.
Figure~{\ref{SurfaceEtchants}} compares two bottom electrodes etched with one liquid and the other.
In the sample shown on the right, the etching process was stopped so soon after the holes in the bottom  were opened, that the edge of the metal is still very sharp.
This is a natural consequence of the isotropic etching through the hole in the polyimide, and can typically be solved by moderate over-etching.

In order to avoid creating the rim in the top electrode completely, one has to protect the metal when etching the bottom electrode.
This can roughly be done in two ways: \emph{galvanic protection} and \emph{electrochemical protection}.
With galvanic protection, the top electrode is hermetically covered by a layer that is not attacked by the etchant.
A few tests have been done using gold as protection layer.
It is suggested that a tin layer may work as well, with the added benefit of lower cost and the possibility to remove the layer afterwards.
However, it turns out to be quite difficult to obtain a good hermeticity, and also the slightest delamination between copper and polyimide leads to a leak through which the etchant can attack the copper, see figure~\ref{GalvanicProtection}.
Due to the dents in the edge of the electrode around the hole, indicated in the figure, these foils tend to spark already at fairly low voltage.
It is therefore difficult to reach sufficient gain for practical applications.
\FIGURE{\includegraphics[width=\textwidth]{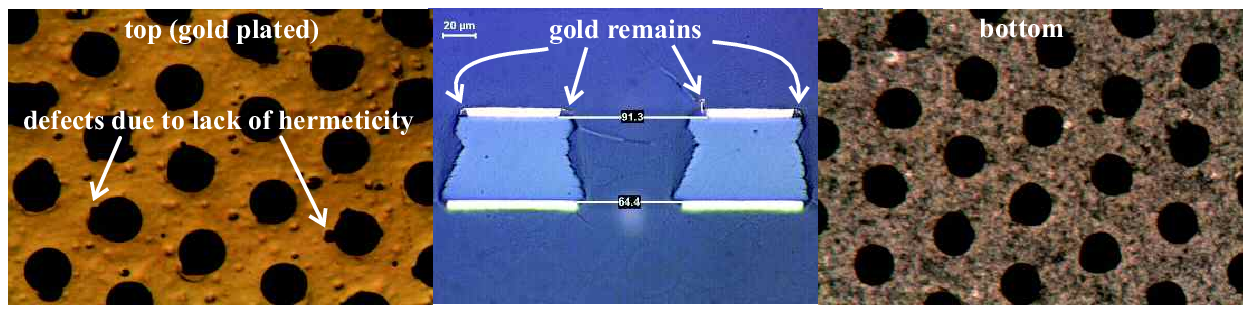}
\caption{Galvanic protection by a very thin gold layer over the top electrode. Due to failing hermeticity of the layer, some copper is etched underneath it.}
\label{GalvanicProtection}}

Electrochemical protection proves to be more successful in avoiding a rim in the top electrode, see figure~\ref{ElectrochemicalProtection}.
It works by giving the top electrode a few volts negative potential with respect to the bottom electrode and the walls of the etching bath.
The top electrode then becomes inert to the reactions that attack the bottom electrode.
This is essentially different from electrolytic etching\footnote{Electrolytic etching has been suggested as a way to etch the holes in the bottom electrode. Results were consistently poor however, and this method was abandoned.}, where a direct current is running from the metal to be etched (i.e. the bottom electrode) to an external electrode.
With electrochemical protection a direct current runs from the external electrode (i.e. the walls of the bath) to the top electrode.
\FIGURE{\includegraphics[width=\textwidth]{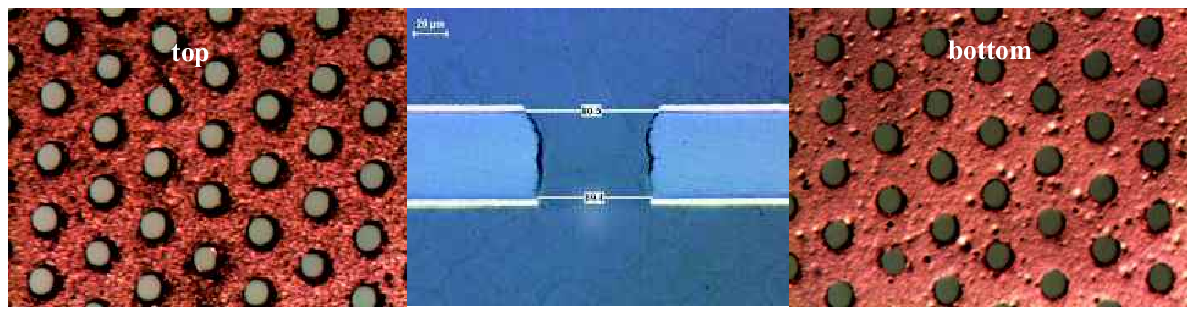}
\caption{Chemical protection of the top electrode. After etching the bottom electrode, a final short polyimide etching gets rid of the bottom rim, and makes the hole almost cylindrical.}
\label{ElectrochemicalProtection}}

The \textsc{gem}s made this way have a gain almost consistent with standard \textsc{gem}s, and are more stable than those shown in figure~\ref{NoProtection}.
However, they seem to be more easily damaged by discharges than other \textsc{gem}s; a single discharge can make a conductive contact ($< 1 \textrm{M}\Omega$) between top and bottom electrodes.
Adding a rim of about $1 \mu$m can solve this issue.

\pagebreak
\section{Simulations}
\FIGURE{\includegraphics[width=\textwidth]{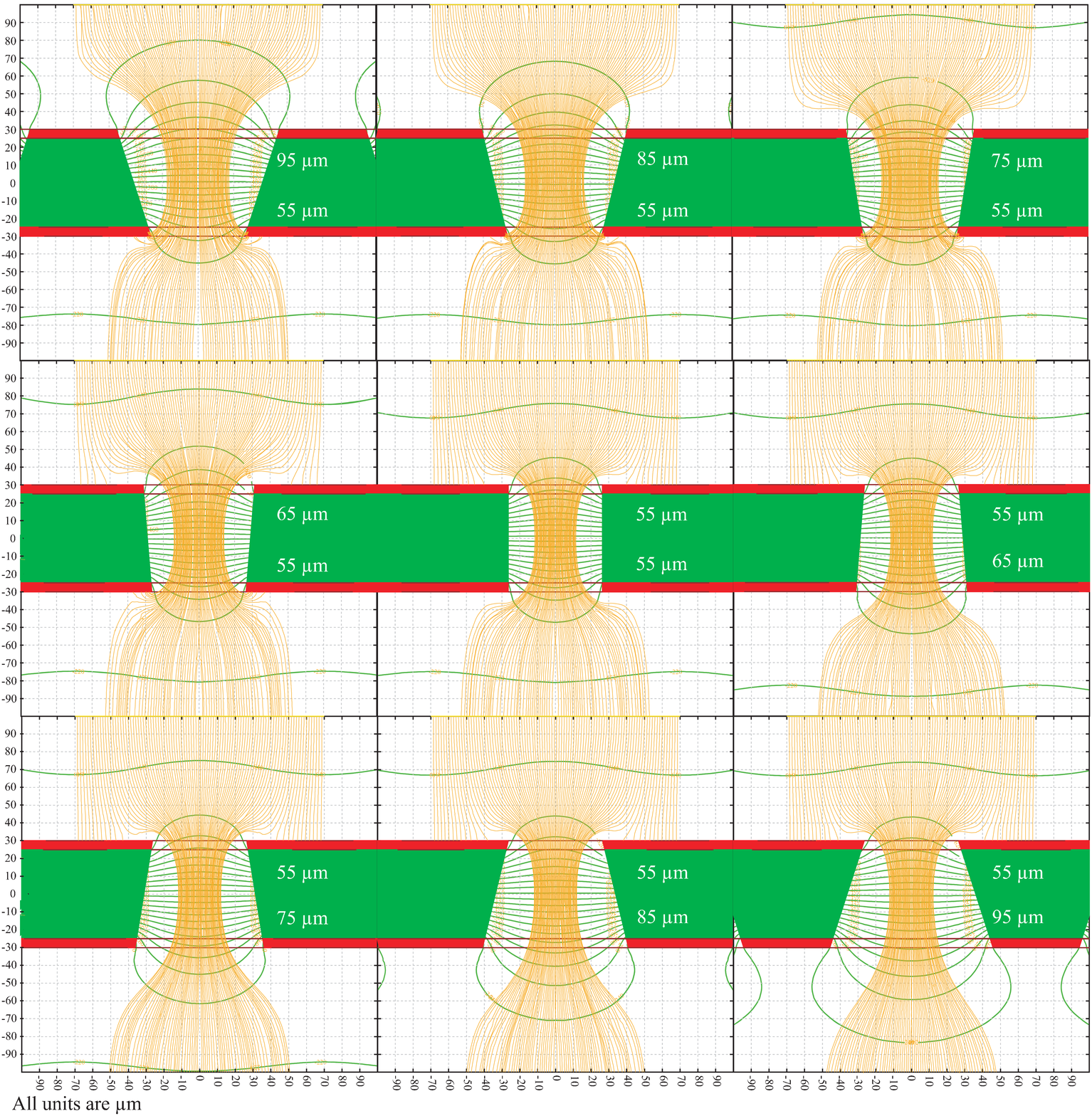}
\caption{Simulated drift lines of electrons in \textsc{gem} holes to study how electron transparency depends on hole shape. Diameters of entrance and exit of each hole are indicated in the frames. Conditions: Ar/CO$_2$ 70/30\%, T$=300$K, p$=1$atm, $\textrm{E}_{\textrm{\scriptsize drift}}=\textrm{E}_{\textrm{\scriptsize induction}}=3$kV/cm, $\Delta\textrm{V}_\textrm{\textsc{\scriptsize gem}}=400$V. Diffusion is not taken into account, neither is secondary ionization, and the primary electrons originate from 100 equidistant points in the drift region.}
\label{NoDiffusion}}
The single-mask technique introduces conical holes, and the various new methods available to make them allows control over their shape.
But it is not yet obvious what would be the optimal shape of a hole, or the optimal orientation (larger diameter towards anode or cathode).
Also, what is regarded as optimal depends on what properties one wants to optimize.
\FIGURE{\includegraphics[width=.4\textwidth]{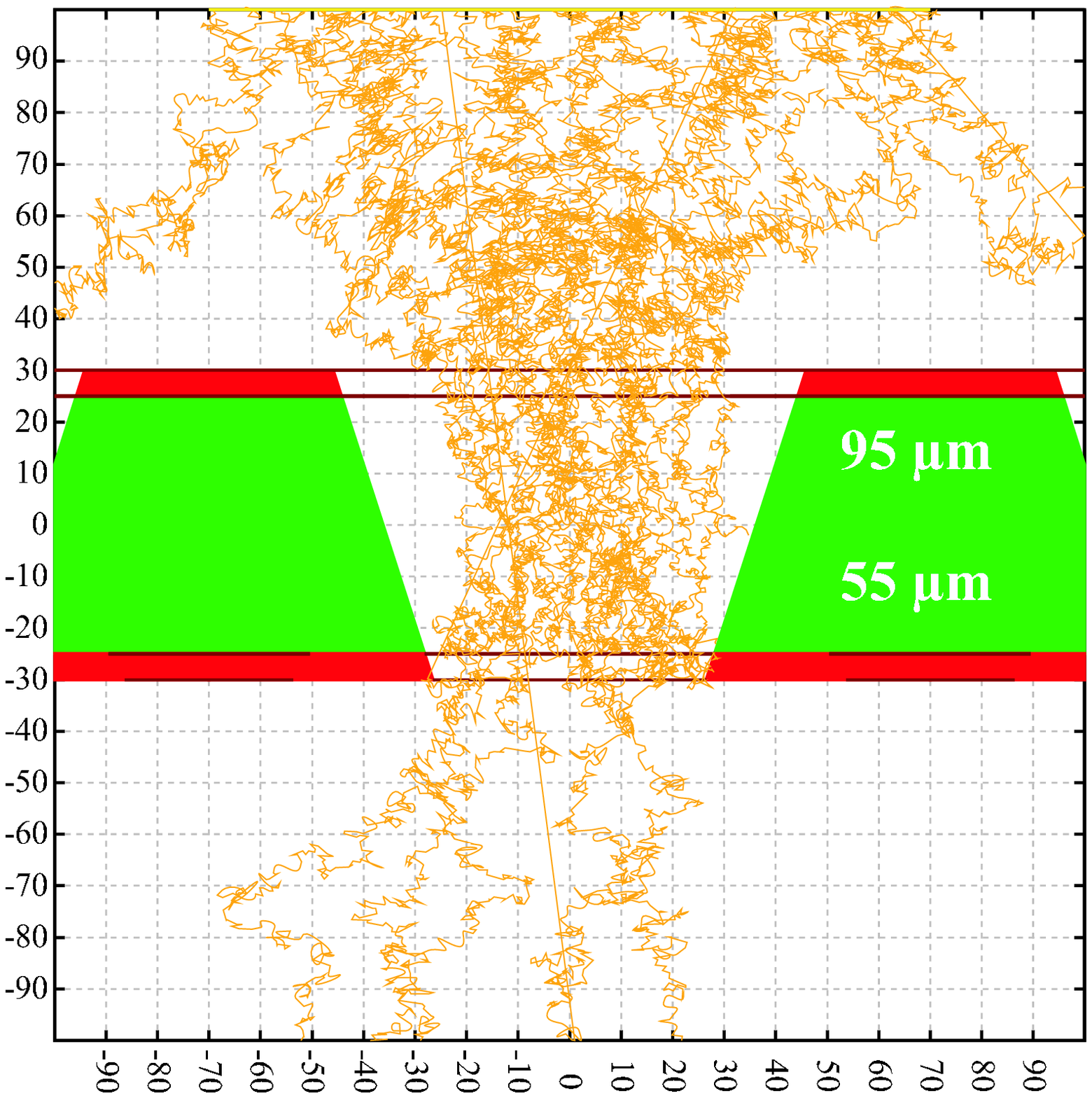}
\caption{Simulation of drifting electrons in a conical \textsc{gem} hole. Diffusion is taken into account, otherwise all conditions are identical to those in figure~\ref{NoDiffusion}.}
\label{Diffusion}}

To gain some understanding of how the shape and orientation of \textsc{gem} holes influence its properties, numerical simulations were done on conical \textsc{gem} holes with various diameters on top and bottom.
Figure~\ref{NoDiffusion} shows drift trajectories of primary electrons in conical holes, assuming no diffusion or secondary ionization.
These drift trajectories are a good approximation of the field lines in and around the hole.
They show how the focusing properties of the field depend on hole geometry.
The field maps were generated with the Ansys finite element method package\footnote{\href{http://www.ansys.com/}{\texttt{http://www.ansys.com/}}}, and the trajectories were calculated by Garfield\footnote{Author: Rob Veenhof (\href{http://garfield.web.cern.ch/garfield/}{\texttt{http://garfield.web.cern.ch/garfield/}})}.

For a more realistic view of drifting electrons diffusion must be taken into account, and Garfield features a very accurate algorithm to simulate diffusion\footnote{This algorithm is called microscopic drift simulation.}.
Each of the geometries shown in figure~\ref{NoDiffusion} is simulated with this algorithm, by liberating 1000 electrons $100 \mu$m above the hole (randomly distributed over the horizontal plane).
Field, gas and environmental conditions are identical to those in figure~\ref{NoDiffusion}.
\FIGURE[l]{\includegraphics[width=.57\textwidth]{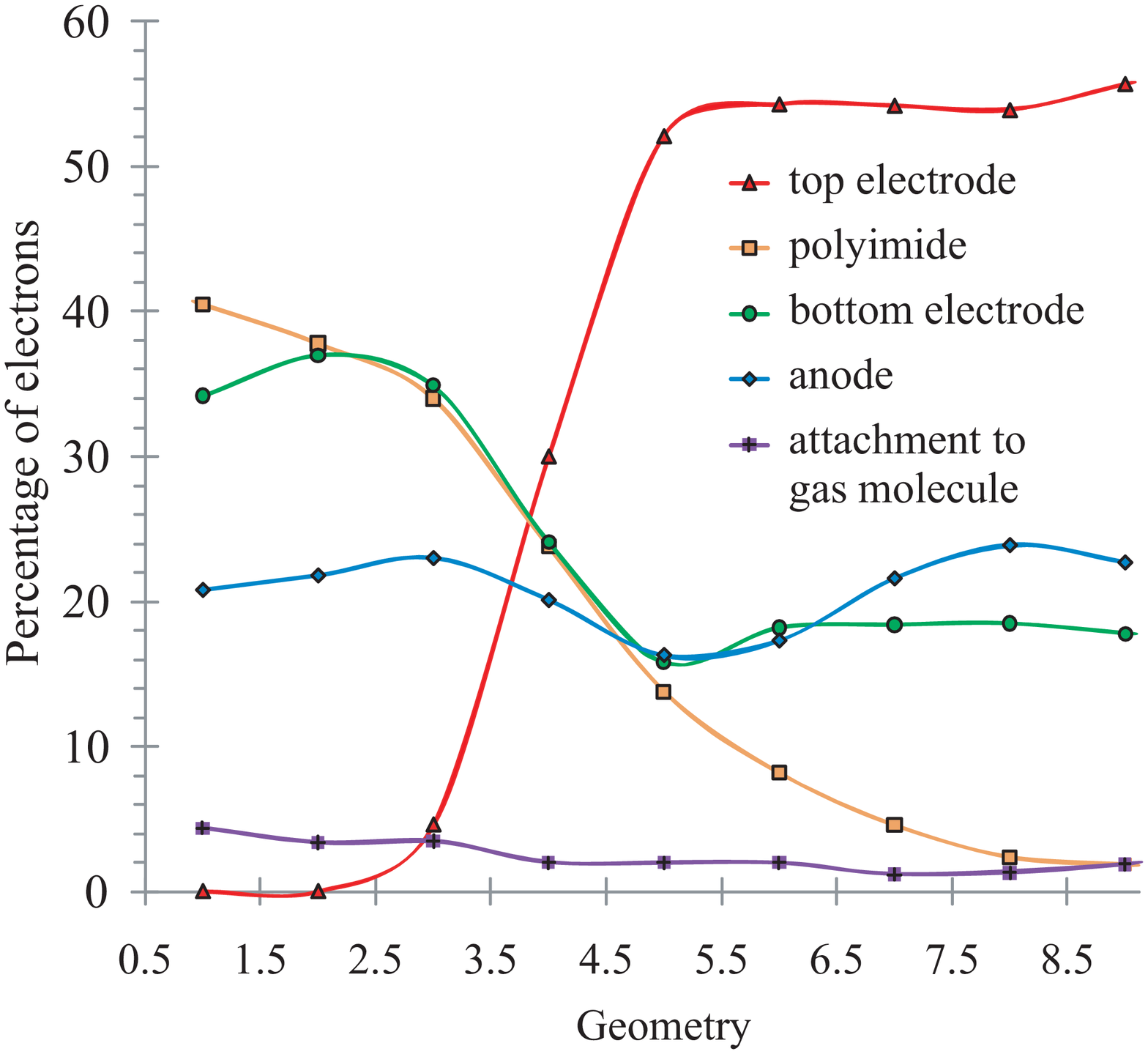}
\caption{Endpoints of electron drift trajectories simulated by the Garfield microscopic tracking algorithm, in the hole geometries shown in figure~\ref{NoDiffusion}. The hole geometry is indicated on the horizontal axis as: $\diameter_{\textrm{\scriptsize top}} - \diameter_{\textrm{\scriptsize bottom}}$, in microns.}
\label{ElectronTransparency}}
It is interesting to see where primary electrons end their trajectory, and how this depends on the hole shape, see figure~\ref{ElectronTransparency}.
The curves most strongly dependent on geometry are the number of electrons ending up at the top elec\-tro\-de and those ending up at the po\-ly\-imi\-de.
This would lead to the con\-clu\-sion that holes wider on top (to\-wards the cathode) are generally good for energy resolution (no primaries lost on top electrode), but worse for time sta\-bi\-li\-ty (stron\-ger dielectric surface charging).\nolinebreak 
This is consistent with our observations with the large prototype, which has holes similar to the left end of the horizontal scale in figure~\ref{ElectronTransparency}.
The over\-all trans\-pa\-ren\-cy (percentage of electrons reaching the readout electrode) seems al\-most independent of the hole shape.
Note that as no avalanche is simulated here, no conclusions can be drawn concerning gain, or maximum gain.
Maximum gain generally depends mainly on high voltage stability, and does not lend itself to simulation.

\section{Production}
\label{production}
The large area prototype generated a lot of interest in large area \textsc{gem}s.
Applications range from forward tracking and triggering in s\textsc{lhc} experiments, to barrel tracking with cylindrical \textsc{gem}s~\cite{CERNLNF}, digital hadronic calorimetry for \textsc{ilc} experiments~\cite{DHCAL}, cosmic muon tomography for detection of high Z materials~\cite{tomography}, time projection chamber readout~\cite{TPC}, and various tracking detectors for nuclear physics experiments.

\FIGURE{\includegraphics[width=.58\textwidth]{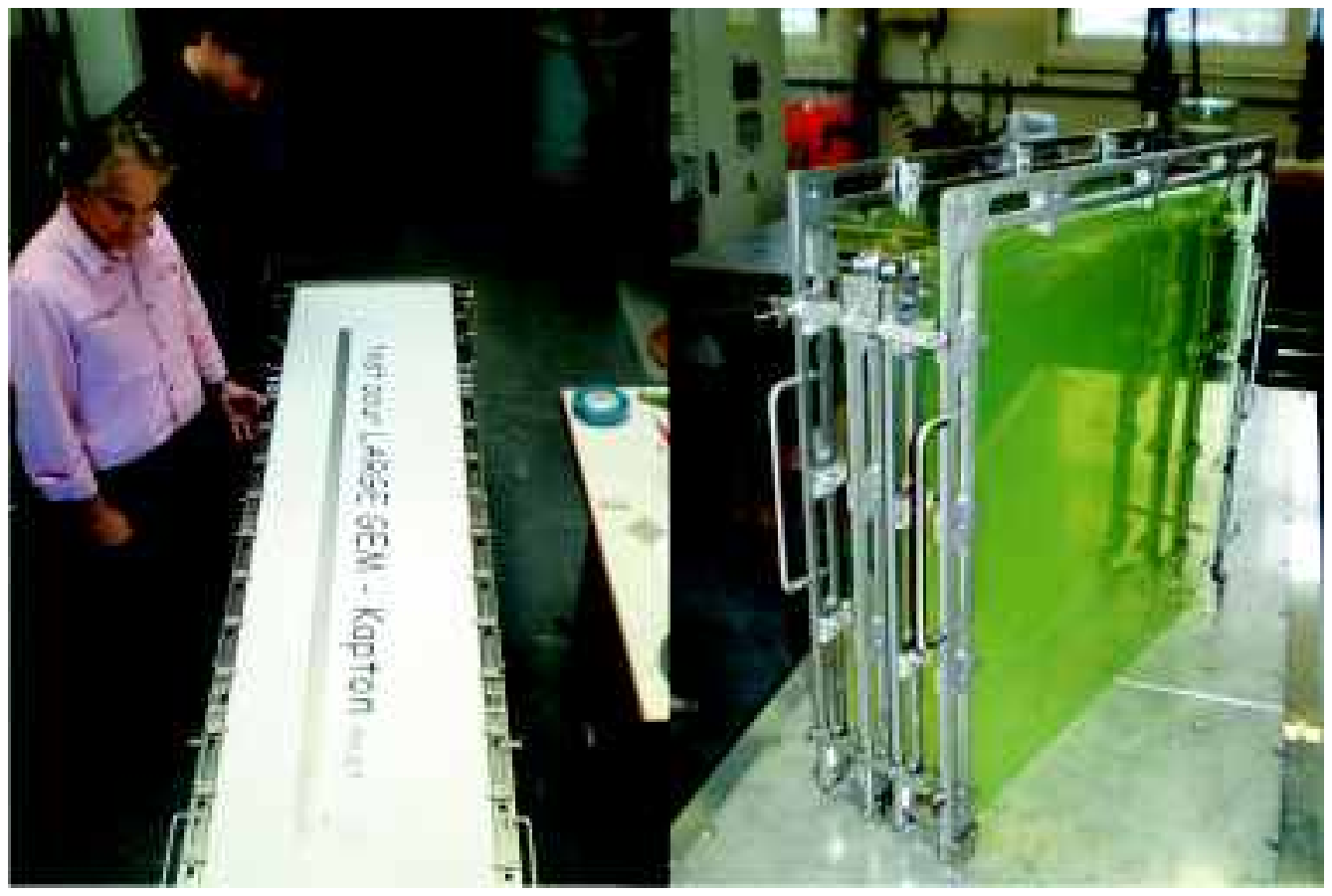}
\caption{Stainless steel portfolio for handling \textsc{gem}s of up to $0.5 \times 2$\,m$^2$ during fabrication. Left in the opened state, as it will be when a foil is mounted or dismounted. Right in the folded state, as it will be when it is moved or immersed in chemistry.}
\label{Portfolio}}
In the fabrication of ever larger area \textsc{gem}s, handling is a growing concern.
Moving large and fragile foils from one bath to another is not trivial, and the baths themselves also have finite dimensions.
A foldable portfolio was designed for this purpose,  see figure~\ref{Portfolio}.
It is made of stainless steel, with no plastic components or lubricants, so that it can be immersed in the etching liquids together with the foil it holds.
This opens the way for fabrication of foils of up to $0.5 \times 2$\,m$^2$ in the \textsc{dem} workshop at \textsc{cern}.

For many of the applications mentioned above, the production volumes foreseen are much larger than for the standard \textsc{gem} detectors made thus far.
Therefore, some industrialization of production is foreseen.
Contrary to standard \textsc{gem}s, single-mask \textsc{gem}s are particularly well suited for mass production.
All steps in the fabrication process can in principle be done with roll-to-roll equipment, due to the absence of involving manual interventions (i.e. the fine alignment of films for each foil separately).
This means that some or all of the steps can be subcontracted to external companies, and the \textsc{gem}s can be transported on their rolls (100 meters long).
The maximum production rate using roll-to-roll equipment is several thousand m$^2$ per month, much greater than required for detector development.
The price per unit area is expected to drop by almost two orders of magnitude compared to standard \textsc{gem}s.

\section{Conclusions}
Single-mask \textsc{gem}s have proven to be a good alternative to standard \textsc{gem}s.
Recent development of the technique gives better control over the shape of holes and the size (or absence) of a rim.
To get a handle on understanding the effect of hole shape on \textsc{gem} performance, simulation studies are performed.

The technique offers attractive advantages for large area and large scale production.
Many applications are foreseen for single-mask \textsc{gem}s, for scientific as well as commercial applications.

\end{document}